\documentclass[preprint,showpacs,preprintnumbers,amsmath,amssymb]{revtex4}

\usepackage{graphicx}
\usepackage{dcolumn}
\usepackage{bm}

\begin{document}


\title{Spectral-based Propagation Schemes for Time-Dependent Quantum Systems with Application to Carbon Nanotubes}

\author{Zuojing Chen}
 \email{zuojing@ecs.umass.edu}
\affiliation{Department of Electrical and Computer Engineering, 
University of Massachusetts, Amherst, Massachusetts 01003, USA}
\author{Eric Polizzi}
\email{polizzi@ecs.umass.edu}
\affiliation{Department of Electrical and Computer Engineering,
 University of Massachusetts, Amherst, Massachusetts 01003, USA}

\date{\today}

\begin{abstract}
Effective modeling and numerical spectral-based propagation schemes are proposed
for addressing the challenges in time-dependent quantum simulations of systems 
ranging from atoms, molecules, and nanostructures to
emerging nanoelectronic devices.
While time-dependent Hamiltonian problems can be formally solved 
by propagating the solutions along tiny simulation time steps, a direct 
numerical treatment is often considered too computationally demanding.   
In this paper, however, we propose to go beyond these limitations
by introducing high-performance numerical propagation schemes
to compute the solution of the time-ordered evolution operator.
In addition to the direct Hamiltonian diagonalizations that 
can be efficiently performed using the  new eigenvalue solver FEAST, 
we have designed  a Gaussian propagation 
scheme and a basis transformed propagation scheme (BTPS) which allow to
reduce considerably the simulation times needed by time intervals.
It is outlined that BTPS offers the best computational efficiency allowing
new perspectives in time-dependent simulations. 
Finally, these numerical schemes are applied  to study 
the AC response of a (5,5) carbon nanotube within a 3D real-space mesh framework.

\end{abstract}

\maketitle

\section{Introduction}\label{secI}

Nowadays, the numerical solution of the time-dependent Schr\"{o}dinger-type equation
is still considered as one of the most challenging problems in quantum simulations of molecules, 
nanostructures and devices. In nanoelectronic applications, in particular,
efficient time dependent simulations have become increasingly important 
for characterizing the electron dynamics under time dependent external perturbations 
such as electromagnetic fields, pulsed lasers, AC signals, particle scattering, etc.
For example, THz frequency responses for carbon nanotube (CNT) 
have recently been observed in \cite{Fu2008} pointing out the need of time-dependent simulations 
capable of going beyond the linear response regime.
Reliable modeling approaches in time domain, however, 
are often limited in term of trade-off between robustness and performances
 \cite{Iitaka1994,Castro2004}.

One approach for solving the time-dependent Schr\"{o}dinger-type equation consists of 
using a partial differential equation representation where one can generally discretize
the time domain using finite difference method. 
The specific techniques include both explicit and implicit schemes, 
with the commonly used Crank-Nicolson scheme \cite{Crank}.
These numerical techniques can be cast as direct approaches, however, they
can end up being numerically expensive in the case of long time quantum simulations
where it is essential to preserve accuracy and robustness.

Another commonly used approach 
consists of solving the integral form of the problem via the numerical treatment of the time-ordered evolution
operator. Two cases can then be generally considered: (i) The Hamiltonian is time-independent; 
(ii) The Hamiltonian is time-dependent. In order to solve this latter,  
one traditional solution consists of dividing the simulation time domain into tiny time
steps $\Delta_t$, such that the Hamiltonian can be considered as time-independent within $\Delta_t$ 
and the system can be solved with techniques similar to case (i).
Case (i) is indeed formally straightforward,  since the problem is then
equivalent to solving the exponential of a Hamiltonian. The most obvious way to
address this numerical problem would be to directly diagonalize the Hamiltonian while
selecting the relevant number of modes (i.e. eigenpairs) needed to accurately expand the solutions.
Spectral decomposition, however, are known to be computationally 
demanding especially for large systems. The mainstream in time dependent simulations
has then been to avoid solving directly the eigenvalue problem and use  approximations most often
 based on 
split operator techniques \cite{Bandrauk1993,Mikhailova1999,Sugino1999}. 
In this paper, however, it is pointed out that direct diagonalizations can now be efficiently performed 
by taking advantage of the recent high-performance  
eigenvalue solver FEAST \cite{Polizzi2009,feast}.   

However, when the Hamiltonian is time-dependent, 
a direct propagation technique still involves solving a very large number of eigenvalue problems
along the whole simulation times. Although, the FEAST solver can be used at each time step to speed up the numerical 
process, we propose here to introduce novel spectral-based propagation schemes that can significantly 
reduce the total number of eigenvalue problems  
while preserving high numerical accuracy and robustness.
The high-efficiency of these techniques  can be achieved, in particular, 
by considering larger $\Delta_t$ and focusing only on obtaining the final states  
at the end of each time interval.

This paper is organized as follows: Section \ref{secII} 
presents the basics of time dependent Schr\"odinger equation within the TDDFT and Kohn-Sham formalism;
Section \ref{secIII} describes the various propagation approach used in this work including a direct scheme, 
 a novel Gaussian quadrature scheme, and an optimized basis transformed propagation scheme (BTPS);
The numerical efficiency of these techniques is finally compared in Section \ref{secIV} 
using simulation results on 3D CNT.


\section{Time-dependence in quantum systems}\label{secII}

In quantum systems, the electrons obey
the time-dependent Schr\"{o}dinger equation:
\begin{equation}\label{tddft}
i\hbar\frac{\partial }{\partial t}\Psi(t)=\hat{H}\Psi(t).
\end{equation}
Using a single electron picture, the Hamiltonian is composed of two terms, one of kinetic origin and another 
describing the interaction of the particle with a local potential which may be time dependent:
\begin{equation}
\hat{H}=\hat{T}+\hat{V}(t)=-\frac{\hbar^2}{2m}\nabla^2+v(\textbf{r},t).\end{equation}
Besides appropriate boundary conditions, 
the time dependent Schr\"{o}dinger equation requires an initial value condition
 $\Psi(t=0)=\Psi_0$ that completely determines the dynamics of the system.
In particular, within  the time dependent 
density functional theory (TDDFT) framework \cite{tddft}, the solutions of the stationary 
Kohn-Sham Schr\"odinger-type equation are taken as initial wave functions and will be propagated 
over time.

TDDFT can be viewed as a reformulation of time dependent quantum mechanics where the basic variable is no longer 
the many body wave function, but the time dependent electron density $n(\textbf{r},t)$.
For a system composed of $N_e$ electrons,  
the electron density can be obtained from the solution of a set of one body equations, 
the so called Kohn Sham equations, that have the same form as Equation (\ref{tddft}) where
$\Psi=\{\psi_1,\psi_2,\hdots,\psi_{N_e}\}$ and with $\psi_j$ solution of:
\begin{equation}
i\hbar\frac{\partial}{\partial t}\psi_j(\textbf{r},t)=\left[-\frac{\hbar^2}{2m}\nabla^2+v_{KS}[n](\textbf{r},t)\right]\psi_j(\textbf{r},t)
\;.\end{equation}
The density of the interacting system can be obtained from the time-dependent Kohn-Sham wave functions
\begin{equation}
n(\textbf{r},t)=\sum_{j=1}^{N_e} |\psi_j(\textbf{r},t)|^2.\end{equation}
The Kohn-Sham potential $v_{KS}$ is a functional of the time-dependent density and it is 
conventionally separated in the following way:
\begin{equation}
\label{eqvks}
v_{KS}[n](\textbf{r},t)=v_{ext}(\textbf{r},t)+v_{H}[n](\textbf{r},t)+v_{xc}[n](\textbf{r},t),
\end{equation}
where the first term represents the external potential, the second term is the Hartree potential which 
accounts for the electrostatic interaction between the electrons, and the last term is defined as the 
exchange-correlation potential which accounts for all the non-trivial many-body effects.
Finally, it should be noted that at $t=0$, the initial wave functions $\Psi_0=\{\psi_1^{(0)},\psi_2^{(0)},\hdots,
\psi_{N_e}^{(0)}\}$ are solutions of the ground state DFT Kohn-Sham stationary equations \cite{dft}:
\begin{equation}
\left[ -\frac{\hbar^2}{2m}+v_{KS}[n](\textbf{r}) \right]\psi_j^{(0)}(\textbf{r})= E_j^{(0)} \psi_j^{(0)}(\textbf{r}).\end{equation}
In a confined system, the Kohn-Sham wave function $\psi_j^{(0)}$ (i.e. $\psi_j(\textbf{r},t=0)$) is associated 
with the eigenvalue $E_j^{(0)}$,  and the ground state many-body density is given by $n(\textbf{r},t=0)$.
Formally, the solution of Equation (\ref{tddft}) can be written as
\begin{equation}\label{oper1}
\Psi(t)=\hat{U}(t,0)\Psi_0=\mathcal{T}\, \exp\left\{  -\frac{i}{\hbar}\int^t_0d\tau \hat{H}(\tau)  \right\}\Psi_0,
\end{equation}
where the evolution operator  $\hat{U}$ is unitary and can be represented using
a time ordered exponential $\mathcal{T} \exp$,
 which is  a non-trivial mathematical object defined in noncommutative algebras.
It is important to note that if the Hamiltonian 
is time-independent, the solution takes a simplified form:
\begin{equation}
\Psi(t)=\exp\left\{-\frac{i}{\hbar} t\hat{H}\right\}\Psi_0.
\label{oper2}
\end{equation}
Unfortunately, this is not the case in most relevant applications which require
the description of the electron dynamics under time dependent external perturbations 
such as electromagnetic fields, 
pulsed lasers, AC signals, particle scattering, etc. In this case, approximations such as perturbation theory or
linear response are commonly used to simplify the computational cost of the time-dependent solutions.
In the following Section \ref{secIII}, however, novel spectral-based propagation schemes
are proposed to address the numerical challenges of 
solving (\ref{oper1}) for the case of large scale 
problems and long simulation time domains. 
In order to ease the description of the numerical techniques, one will
consider only non-interacting systems (i.e. the potential $v$ is time dependent but it is not a functional of $n$).
The natural extension of these numerical schemes to solving non-linear problems will be discussed in Section \ref{secV}.


\section{Spectral-based Propagation Schemes}\label{secIII}

\subsection{Direct propagation approach}\label{secIIIa}

In practice,
  intermediate physical solutions 
are computed in addition 
to the final solution $\Psi(t)$,
in order  
to describe the evolution of the system over $[0,t]$.
This can be accomplished by dividing  $[0,t]$ into smaller time intervals since using the intrinsic properties
of the evolution operator, one can apply the following decomposition:
\begin{equation}\label{unit}
\hat{U}(t,0)= 
\hat{U}(t_n,t_{n-1})\hat{U}(t_{n-1},t_{n-2})\dots\hat{U}(t_2,t_1)\hat{U}(t_1,t_0),
\end{equation}
where we consider $n-1$ intermediate interval time steps with $t_0=0$, and $t_n=t$.
 Most often a constant time step $\Delta_t$ is used 
 and one has to deal now with the problem of performing many shorter time propagation of the solutions 
along the whole interval $[0,t]$:
\begin{equation}\label{prop1}
\Psi(t+\Delta_t)=\mathcal{T} \exp\left\{  -\frac{i}{\hbar}\int^{t+\Delta_t}_t d\tau \hat{H}(\tau)  \right\}\Psi(t).
\end{equation}
Additionally, if $\Delta_t$ is chosen very small, it is reasonable to 
consider $\hat{H}(\tau)$ constant within the time interval $[t,t+\Delta_t]$ leading to
\begin{equation}\label{bf1}
\Psi(t+\Delta_t)=\hat{U}(t+\Delta_t,t)\Psi(t)=\exp\left\{ -\frac{i}{\hbar}\Delta_t \hat{H}(t)   \right\}\Psi(t),\end{equation}
which is also equivalent to solving the time independent problem (\ref{oper2}).

Denoting $\mathbf{H}$ the $N\times N$ Hamiltonian matrix obtained after the discretization of $\hat{H}$ 
at a given time $t$ and where $N$ could represent the number of basis functions (or number of nodes 
using real-space mesh techniques),  $\mathbf{H}$ can then be diagonalized as follows:
\begin{equation}\label{eig}
\mathbf{D}=\mathbf{P}^{T} \mathbf{H}\mathbf{P},
\end{equation}
where the columns of the matrix $\mathbf{P=\{p_1,p_2,\dots,p_M\}}$ represent the eigenvectors of 
$\mathbf{H}$ associated with the $M$ lowest eigenvalues 
regrouped within the diagonal matrix $\mathbf{D}=\mathrm{\{d_1,d_2,\dots,d_M\}}$.
If the discretization is performed using non-orthogonal basis functions (e.g. finite element basis functions 
in real-space), the eigenvalue 
problem that needs to be solved takes the generalized form:
\begin{equation}\label{eig1}
\mathbf{H}\mathbf{p_i}=\mathrm{d_i}\mathbf{S}\mathbf{p_i},
\end{equation}
where $\mathbf {S}$ is a symmetric positive-definite matrix. Here, we consider that the computed eigenvectors 
$\mathbf{P}$ are  $\mathbf{S}$-orthonormal i.e.  $\mathbf{P}^{T}\mathbf{S}\mathbf{P}=\mathbf{I}$.
In order to evaluate the exponential of  the Hamiltonian in (\ref{bf1}), it is now possible to 
perform a spectral decomposition where the exponential acts only on the eigenvalue matrix $\mathbf{D}$. 
One can show that the resulting matrix form of time propagation equation is given by:
\begin{equation}
\mathbf{\Psi}(t+\Delta_t)=\mathbf{P} \,\exp\left\{    -\frac{i}{\hbar}\Delta_t
 \mathbf{D} \right\} \, \mathbf{P}^T \mathbf {S}\mathbf{\Psi}(t),\end{equation}
which is exact if  $M=N$. We also note that $\mathbf{\Psi}^{T}\mathbf{S}\mathbf{\Psi}=\mathbf{I}$
with $\mathbf{\Psi}=\{\mbox{\boldmath $\psi$}_1,\mbox{\boldmath $\psi$}_2,\hdots,\mbox{\boldmath $\psi$}_{N_e}\}$.
 In practice, it is reasonable to obtain very accurate spectral approximations even if one selects 
the number of lowest eigenvalues $M$ much smaller than the size of the system $N$ but greater than $N_e$
(i.e. $N_e<M<<N$).
The $N_e$ propagating states $\mathbf{\Psi}$ are indeed low energy states. 
In this case,  $\mathbf{\Psi}$, $\mathbf{P}$, $\mathbf{D}$, $\mathbf{S}$ 
are respectively matrices of size ${N\times N_e}$,  ${N\times M}$,  ${M\times M}$, and  ${N\times N}$. 
Using the property (\ref{unit}), the solution $\mathbf{\Psi}(t)$ can finally be obtained in function of $\mathbf{\Psi}_0$:
\begin{equation}\label{direct}
\mathbf{\Psi}(t)=\mathcal{T}
\left\{
 \prod_i\left[ \mathbf{P}_i \exp\left(-\frac{i}{\hbar}\Delta_t 
\mathbf{D}_i\right) \mathbf{P}^T_i \mathbf{S} \right]   \right\}
 \mathbf{\Psi}_0,\end{equation}
where $\mathbf{D}_i=\mathbf{P}_i^T\mathbf{H}(t_i){\mathbf{P}_i}$, and the time ordering is defined by
\begin{equation}\label{to}
\mathcal{T}\left\{\prod_i A(t_i)\right\}=A(t_N)\dots A(t_2)A(t_1).
\end{equation}

Therefore, the direct approach 
may require solving hundred to thousand of eigenvalue problems all along the time domain 
(one eigenvalue problem for each time step). For large-scale systems 
(e.g. where hundreds of atoms are taken into consideration), this approach is often
considered not applicable as accurate eigensolutions for thousands
 of eigenpairs are computationally demanding and algorithmically challenging.
Recently, however, the new density-matrix based algorithm FEAST \cite{Polizzi2009} 
has been proposed to overcome these difficulties,
and provide both accuracy, robustness and performance scalability for solving eigenvalue problems.
FEAST is a general purpose algorithm which takes its inspiration from the density-matrix representation
and contour integration technique in quantum mechanics.  Its main computational
tasks consist of solving very few independent linear systems (typically eight linear systems
that can be solved simultaneously in parallel)
and one reduced dense eigenvalue problem orders of magnitude smaller than the original one (of size $M\times M$, 
in the present case). FEAST is also ideally suited for the direct propagation scheme presented here, 
since it can take advantage of the subspace computed at a given time step $i$ as initial
guess for the next time step ${i+1}$ in order to speed-up the numerical convergence.
Ultimately, one can show that if enough parallel computing power is available at hand,
the main computational cost of FEAST  for solving the eigenvalue problem
can be  reduced to solving only one linear system. 
In contrast to the proposed FEAST approach for the direct spectral-based propagation scheme,
 PDE-based numerical schemes for solving (\ref{oper1}) (such as Crank-Nicolson) 
would require solving successively a large number of linear systems by tiny time intervals $\Delta_t$.
A high-performance implementation of the FEAST algorithm can be found in \cite{feast}.

\subsection{Gaussian propagation scheme}\label{secIIIb}

The direct approach has a sounded physical interpretation as it corresponds to a step by step 
propagation of the solution over time. Its major drawback, however, is that it involves a very large number of time 
steps from initial to final simulation times. 
Let us now consider the case of a much greater time interval $\Delta_t$ which may correspond, for instance, to a given
period of an external time-dependent perturbation. The solution can then be obtained by performing first
a numerical integration for the integral on the Hamiltonian in (\ref{prop1}):
\begin{equation}\label{oper111}
\Psi(t+\Delta_t)=\mathcal{T}\exp\left\{  -\frac{i}{\hbar}\xi\sum^p_{i=1}\omega_i \hat {H}(t_i) \right\}\Psi(t)
\end{equation}
where $\omega_i$ and $\xi$ are integration constant, and $p$ is the number of quadrature points.  
 If the quadrature points are close enough, one can additionally assume that the
anti-commutation error between Hamiltonians can be ignored, and the exponential can then be decomposed 
into a product of exponentials for each time step
 \begin{equation}\label{oper1111}
\Psi(t+\Delta_t)=
\mathcal{T}\left\{  \prod_{i=1}^p \exp\left\{-\frac{i}{\hbar}\xi\omega_i \hat {H}(t_i)\right\} \right\}\Psi(t).
\end{equation}
The number of exponentials to evaluate along $[0,t]$ is then equal to the number of time interval $n$ multiply
by the number of quadrature points $p$ by intervals. 
After discretization and spectral decomposition of each Hamiltonian as presented in Section \ref{secIIIa}, it comes:
\begin{equation}
\mathbf{\Psi}(t+\Delta_t)=\mathcal{T}\left\{ \prod_{i=1}^p\left[ \mathbf{P}_i \exp\left(-\frac{i}{\hbar}\xi\omega_i
\mathbf{D}_i\right) \mathbf{P}^T_i \mathbf{S} \right]   \right\} \mathbf{\Psi}(t).\end{equation}

We note from the derivation of equations (\ref{oper111}) and (\ref{oper1111}) that
 two numerical errors are respectively involved:
 (i) an integration error and (ii) an error on the anti-commutation resulting from 
 the decomposition of the exponential. 
The direct propagation scheme presented in Section \ref{secIIIa} can be derived by assuming a rectangle quadrature
rule with $\omega_i=1$ and using $\xi=\delta_t$, $t_i=t+(i-1)*\delta_t$ where 
$\delta_t\equiv \Delta_t/(p-1)$ corresponds to a very small 
time step within intervals. Using this less conventional derivation, 
it is possible to point out that a drastic reduction of the integration error
 (\ref{oper111}) could be obtained by considering higher-order quadrature scheme such as Gaussian quadrature. 
A $p$-point Gaussian quadrature rule is a numerical integration constructed to yield
 an exact result for polynomials of degree $2p-1$ by a suitable
 choice of the points $t_i$ and (Gauss-Legendre) weights $\omega_i$. 
We associate the quadrature points $t_i$ at the Gauss node $x_i$
using $t_i=\frac{\Delta_t}{2}x_i+\frac{2t+\Delta_t}{2}$, also we note $\xi=\Delta_t/2$.
Gaussian quadrature can use relatively much fewer points than
low-order quadrature rule such as rectangle, to yield a high order approximation of the integral of a function.
Interestingly, a large number of numerical experiments
have shown that accurate exponential decomposition (\ref{oper1111}) can still be obtained 
while using much fewer points $p$ or larger spacing within intervals
(even in the presence of strong perturbations). The same experiments have demonstrated, however, 
that increasing the accuracy  
on the numerical integration plays an important role 
for obtaining the correct solutions. 
We will show in Section \ref{secIV} that the Gaussian  quadrature scheme 
provides a good combination of reduced computational consumption 
and high numerical accuracy. The number of exponentials that needs to be evaluated by time intervals
can indeed be reduced by a factor $3$ to $4$ as compared to the direct approach. Using the Gaussian quadrature, 
however, the intermediate solutions obtained by propagating the solutions within the intervals  have no
 physical meaning.  Since the quadrature weights $\omega_i$ are different for each Gauss node, 
 the solutions can indeed only be known at the beginning and at the end of each interval.

\subsection{Basis transform propagation scheme (BTPS) }\label{secIIIc}

By focusing on obtaining only the final states at the end of each time interval $\Delta_t$, the 
Gaussian propagation schemes can considerably reduce the number of eigenvalue problems that needs to be solved.
Yet, solving large scale eigenvalue problems is still  the most computational consuming part in our simulation.
Here, we introduce a basis transform propagation scheme (BTPS) which can help in reducing
not only the number of eigenvalue problems to solve by intervals, but also the size of each eigenvalue problem.

Before performing the decomposition of the exponential, let us first consider the discretization 
of equation (\ref{oper111}):
\begin{equation}\label{doper111}
\mathbf{\Psi}(t+\Delta_t)=\mathcal{T}\exp\left\{  -\frac{i}{\hbar}\xi\sum^p_{i=1}\omega_i \mathbf{H}_i 
\right\}\mathbf{\Psi}(t),
\end{equation}
where $\mathbf{H}_i\equiv \mathbf{H}(t_i)$.
We propose then to project the ``pseudo-Hamiltonians'' $\omega_i\mathbf{H}_i$ onto a common eigen-subspace $\mathbf{P}$ 
constructed from the result of the quadrature sum  $\sum^p_{i=1}\omega_i\mathbf{H}_i$. Denoting $\mathbf{H}$
this new ``global Hamiltonian'' which takes the overall contribution of the time-dependent perturbation over a given
time interval $\Delta_t$ (however $\mathbf{H}$ has no physical meaning), one can perform the diagonalization 
defined in (\ref{eig}) i.e.
\begin{equation}\label{btps1}
\mathbf{D}=\mathbf{P}^T\mathbf{H}\mathbf{P}\equiv \mathbf{P}^T\left\{\sum_{i=1}^p\omega_i\mathbf{H}_i\right\}\mathbf{P},
\end{equation}
where the  subspace $\mathbf{P}$ and eigenvalues $\mathbf{D}$ are obtained solving
the eigenvalue problem in (\ref{eig1}). Denoting  $\mathbf{h}_i$ 
the projection of the pseudo-Hamiltonians at a given quadrature point such that 
\begin{equation}\label{btps2}
\mathbf{h}_i=\mathbf{P}^T(\omega_i \mathbf {H}_i) \mathbf{P},
\end{equation}
it comes with (\ref{btps1})
\begin{equation}\label{btps3}
\mathbf{D}=\sum_{i=1}^p \mathbf{h}_i,
\end{equation}
where $\mathbf{h}_i$ are $M\times M$ dense matrices while their summation $\mathbf{D}$ is diagonal.
Using a spectral decomposition on the time-ordered exponential for the global Hamiltonian, 
the expression (\ref{doper111}) can now take the following form:
\begin{equation}\label{2doper111}
\mathbf{\Psi}(t+\Delta_t)=\mathbf{P}\mathcal{T}\exp\left\{  -\frac{i}{\hbar}\xi\sum^p_{i=1}\mathbf{h}_i 
\right\}\mathbf{P}^T\mathbf{S}\mathbf{\Psi}(t),
\end{equation}
where $\mathbf{P}$ has been moved outside the expression of the time-ordered exponential since
it is a common subspace for all different quadrature time points within $[t,t+\Delta_t]$.
It is then important to note that one cannot replace directly the sum of the projected Hamiltonian 
in the exponential by $\mathbf{D}$ (\ref{btps3}), as the time-ordered operator can only be resolved using a product
 of functions taken at different times (\ref{to}). Similarly to the derivation of (\ref{oper1111}),
we assume that the quadrature points are close enough that one can decompose the exponential into a product
of exponentials for each time step:
\begin{equation}\label{doper1111}
\mathbf{\Psi}(t+\Delta_t)=\mathbf{P}\mathcal{T}\left\{\prod_{i=1}^p\exp\left\{  -\frac{i}{\hbar}\xi\mathbf{h}_i 
\right\}\right\}\mathbf{P}^T\mathbf{S}\mathbf{\Psi}(t),
\end{equation}
Thereafter, one can perform the diagonalization of each $\mathbf{h}_i$ matrices 
\begin{equation}\label{btps4}
\mathbf{h}_i=\mathbf{q}^T_i\mathbf{e}_i\mathbf{q}_i,
\end{equation}
where the $M\times M$ matrices $\mathbf{q}_i$ represent the eigenvectors of 
$\mathbf{h}_i$, and the diagonal matrices $\mathbf{e}_i$ 
 regroup the associated $M$ eigenvalues. In contrast to the generalized eigenvalue problem on the global Hamiltonian above,
 these eigenvalue problems are standard  i.e.: $\mathbf{h}\mathbf{q}=\mathbf{q}\mathbf{e}$.
Finally using a spectral decomposition on the exponentials in (\ref{doper1111}), we obtain:
\begin{equation}\label{2doper1111}
\mathbf{\Psi}(t+\Delta_t)=\mathbf{P}\mathcal{T}\left\{\prod_{i=1}^p\left[\mathbf{q_i}\exp\left(-\frac{i}{\hbar}\xi\mathbf{e}_i \right)\mathbf{q_i}^T\right]\right\}\mathbf{P}^T\mathbf{S}\mathbf{\Psi}(t).
\end{equation}
In contrast to the propagation schemes presented in Sections \ref{secIIIa} and \ref{secIIIb}
 which required solving $n*p$ large
scale eigenvalue problems of size $N\times N$ along $[0,t]$ ($n$ number of intervals $\Delta_t$, and $p$ number
 of quadrature points), the BTPS approach consists of solving only one $N\times N$ eigenvalue problem 
by time intervals (\ref{btps1}), 
and $n*p$ reduced dense eigenvalue problems of size $M\times M$ (\ref{btps4}).

The values of $\xi$, $t_i$ and $\omega_i$  have been specified
in Section \ref{secIIIb} in the case of a rectangle and Gaussian quadrature rules. 
We recall that 
the Gaussian scheme has been proposed to reduce the number of quadrature points $p$ (i.e. number of exponential
to evaluate)  within interval.
From equation (\ref{2doper111}), however, one can see that the numerical integration is here actually 
performed on the reduced projected Hamiltonians, and this basis transformation is expected
to significantly decrease the integration error. As a result, we have found that 
the rectangle quadrature rule is likely to provide the same accuracy than the Gaussian scheme
 using the same number of quadrature points $p$. Since this latter cannot be reduced below a certain threshold in order
to limit the error arising from the decomposition on the exponentials (\ref{doper1111}), 
a high-order integration scheme such as Gauss quadrature becomes then obsolete using BTPS.
Similarly the Gaussian scheme, however, intermediate solutions that can be obtained
by introducing $\mathbf{I}=\mathbf{P^TSP}$ within the product expression in (\ref{2doper1111}),
would end up having no physical meaning. Indeed, one can demonstrate that using BTPS 
the (S)-orthonormalization of the wave functions is only satisfied between the solutions taken at 
the beginning and at the end of each interval.

\section{Simulation results}\label{secIV}

We propose to perform
time-dependent 3D simulations of an isolated carbon nanotube (CNT)
which is sandwiched 
between two electrodes  producing
a AC voltage at the THz frequency. In this model, 
 the charge transfer at the contacts is not considered (i.e. open systems and transport problems are not considered). 
Here, our model uses a local empirical pseudopotential approach, real-space mesh techniques for discretization 
(finite element method), and a time-dependent version of the atomistic mode approach 
described in \cite{Zhang2008,Zhang2009}. Moreover, the empirical pseudopotential $U_{eps}$ is supposed time-independent 
the total atomistic potential can then be decomposed as follows:
\begin{subequations}
\begin{equation}
U(x,y,z,t)=U_{eps}(x,y,z)+U_{ext}(x)\sin(\omega t),\end{equation}
\begin{equation}
U_{ext}(x)=\frac{2x-L}{L}U_0,\end{equation}
\end{subequations}
where $x$ is the longitudinal direction of the tube, $L$ represents the distance between contacts ($x\in[0,L]$), 
$\omega=2\pi f$, with $f$ the corresponding frequency of the AC signal and $U_0$ its amplitude. 
The time-dependent external potential applied to the CNT, maintains then zero in the middle of the tube
but oscillates at both ends alternatively to the $\pm U_0$ values.

Here, the system under consideration is a $6$-unit cell of a (5,5) CNT where the contact-contact distance 
is set at $1.98nm$. This CNT is composed by $120$ atoms, and
using the empirical pseudopotential for Carbon atom 
proposed in \cite{Mayer2004}, one can expect to capture all the $60$ electrons (without spin-dependence) contained in the 
$\pi_z$ orbitals. Using the notations defined in Section \ref{secIII}, 
it comes $N_e=60$, and  $M=120$ ($M$ being the number of 
eigenpairs used for the spectral decomposition and chosen here as twice the 
number of electrons $N_e$). The finite element
discretization of the full 3D system gives rise to sparse matrices of size $N\sim 500*10^3$, while
this size can be drastically reduced in our case using a mode approach in real-space to $N\sim 15*10^3$. 
For the external perturbation (AC signal), we consider $U_0=5eV$ and $f=200THz$.
In our simulation, the solution wave functions 
 $\mathbf{\Psi}=\{\mbox{\boldmath $\psi$}_1,\mbox{\boldmath $\psi$}_2,\hdots,\mbox{\boldmath $\psi$}_{N_e}\}$
will be propagated from $t=0$ to
$t=8T$ where \mbox{$T=1/f=5*10^{-15}s$} denotes the period of the AC signal.
The time interval used is $\Delta_t=T$ where we have found that $p=120$ integration points by intervals 
are sufficient for the direct propagation scheme to provide an accurate reference solution. 
Let us recall that  direct scheme allows to capture the evolution of all the wave functions
 at each tiny time step $\Delta_t/p$,
while the Gaussian and BTPS approaches are expected to produce only accurate results at each time interval 
$\Delta_t$ which length has been arbitrarily chosen very large here to point out the robustness of the approaches.

\begin{figure}
  \includegraphics[width= \columnwidth]{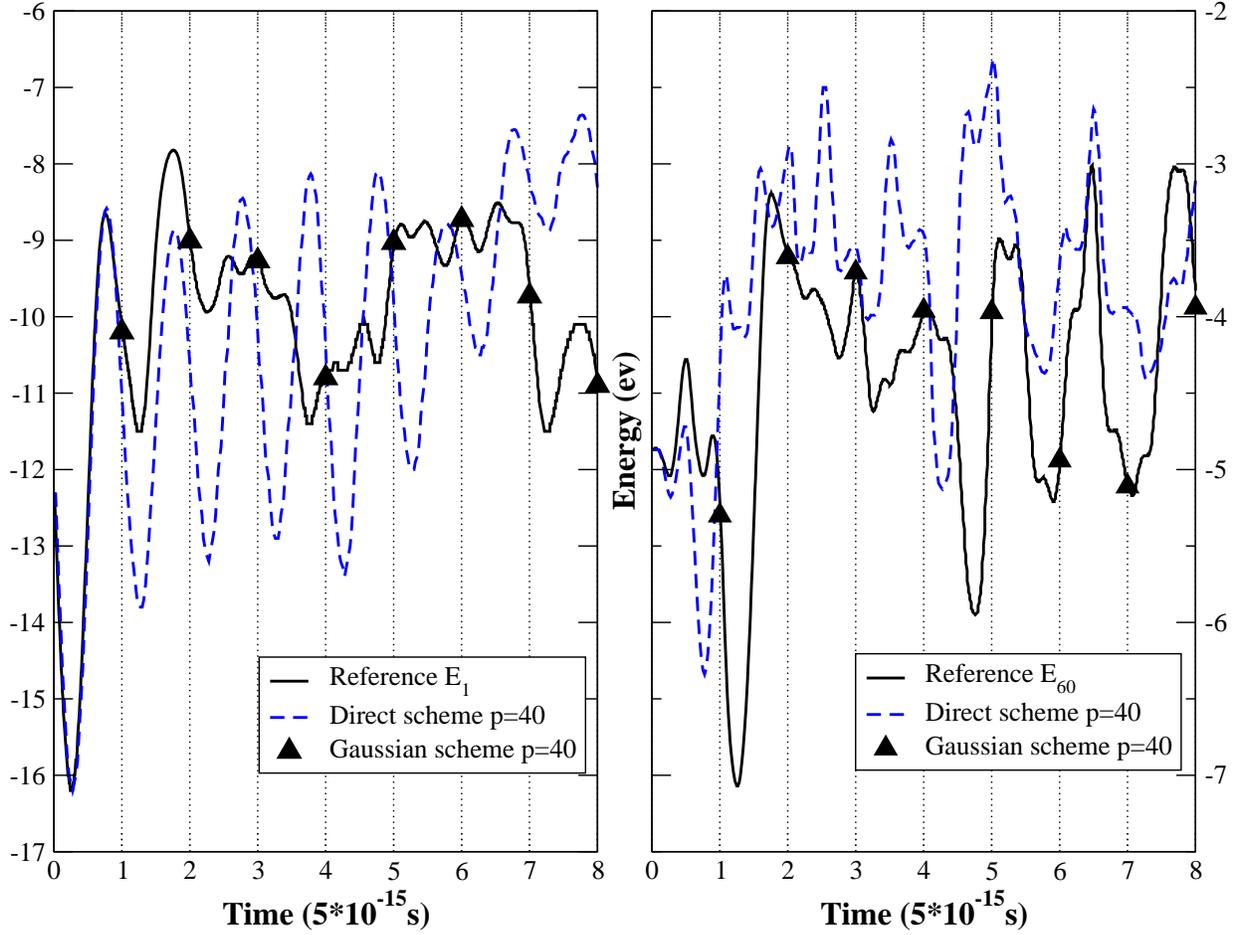}
  \caption{Evolution of the energy expectation of the first level (on the left) and the HOMO level (on the right) along $8$ 
time periods of the AC signal.
  The straight lines represent the reference solutions calculated using the direct scheme with $p=120$ rectangle
quadrature points within intervals. The results for the energy evolution using a direct scheme $p=40$, represented
using dashed lines, diverges after few time steps, while the same number of points did suffice for 
the Gaussian scheme to accurately capture the 
solutions at the end of each time period (represented by filled triangle symbols).
  }
  \label{fig1}
\end{figure}

\begin{figure}
  \centering
  \includegraphics[width= \columnwidth]{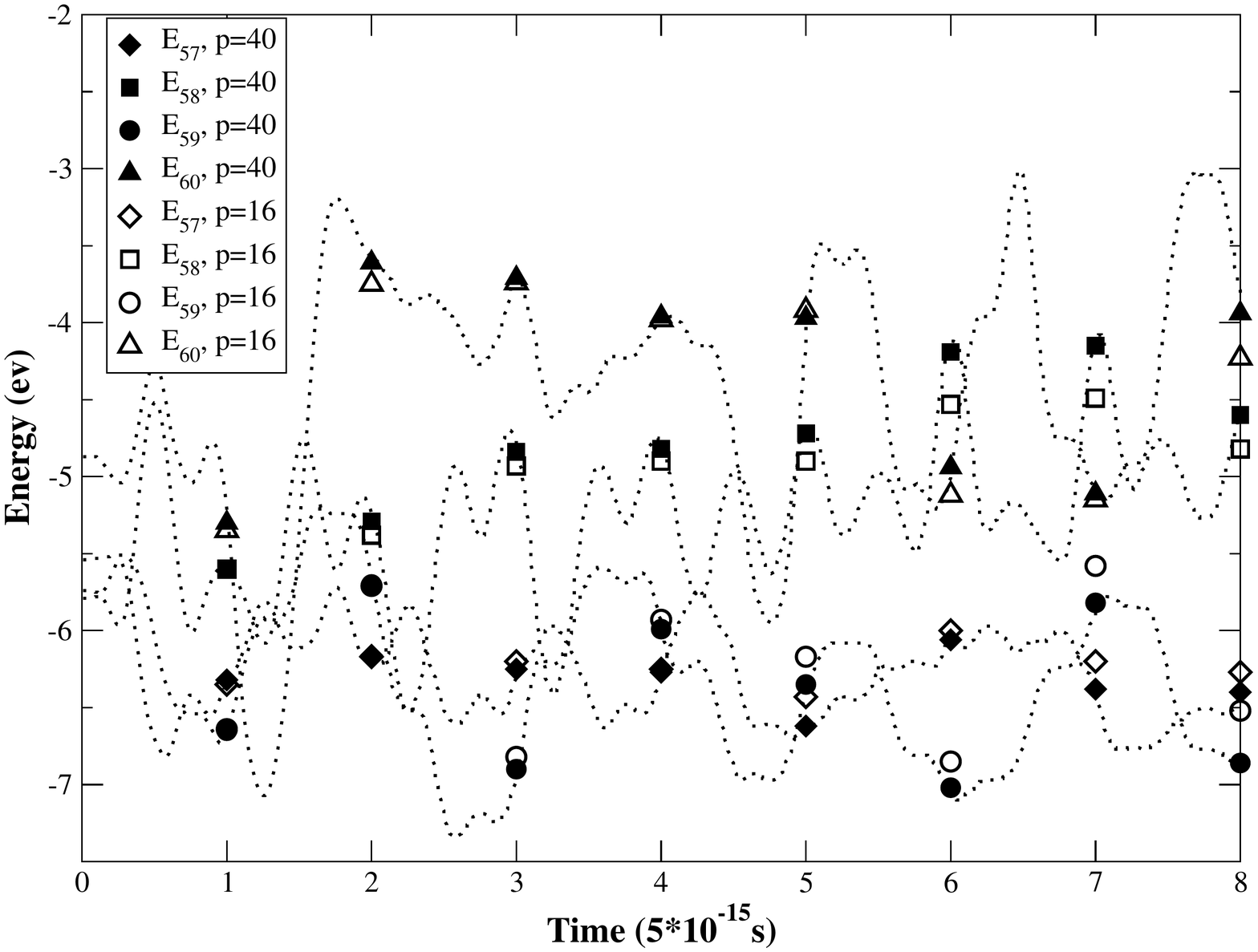}
  \caption{Evolution of the energy expectation the levels $57$ to $60$. The filled symbols are obtained  
using $p=40$ Gaussian propagation scheme, and match exactly the reference solutions (dashed lines) 
at the end of each time period.
In contrast, the solutions obtained using the $p=16$ Gaussian scheme, represented by the unfilled symbols, 
suffer now from an unsuitable approximation on the decomposition of the exponential (\ref{oper1111}) due to 
an increase in distance between integration points.}
  \label{fig2}
\end{figure}

In order to examine the relative accuracy and robustness of the numerical techniques presented in this article,
we propose to calculate the energy evolution of the wave functions i.e.
\begin{equation}\label{energy_form1}
E_j(t)=\langle\mbox{\boldmath $\psi$}_j(t)|\mathbf{H}(t)|\mbox{\boldmath $\psi$}_j(t)\rangle, \quad j=1,\dots,N_e.
\end{equation}
Let us first discuss the integration error which is introduced in our model in (\ref{oper111}). 
As an example, Figure \ref{fig1} compares the direct and Gaussian propagation schemes by
representing the energy evolution of both the first energy level $E_1$ and the HOMO level of the system 
$E_{60}$. Using a direct scheme with $p=40$ integration points for performing the rectangle quadrature 
between time intervals and as compared to the reference curves obtained for $p=120$, 
one can see that the results start diverging after few time steps.
In contrast to the direct scheme, accurate results can still be obtained at the end of each time period using $p=40$ and 
higher order integration technique such as our proposed Gaussian scheme.
In practice, Gauss quadrature could accurately perform the numerical integration (\ref{oper111})
using much fewer integration points ($p<40$), however,  
 this would also increase the anti-commutation error arising from the decomposition 
of the exponential in (\ref{oper1111}) as discussed in Section \ref{secIIIb} 
and shown in Figure \ref{fig2}.

\begin{figure}
  \centering
  \includegraphics[width= \columnwidth]{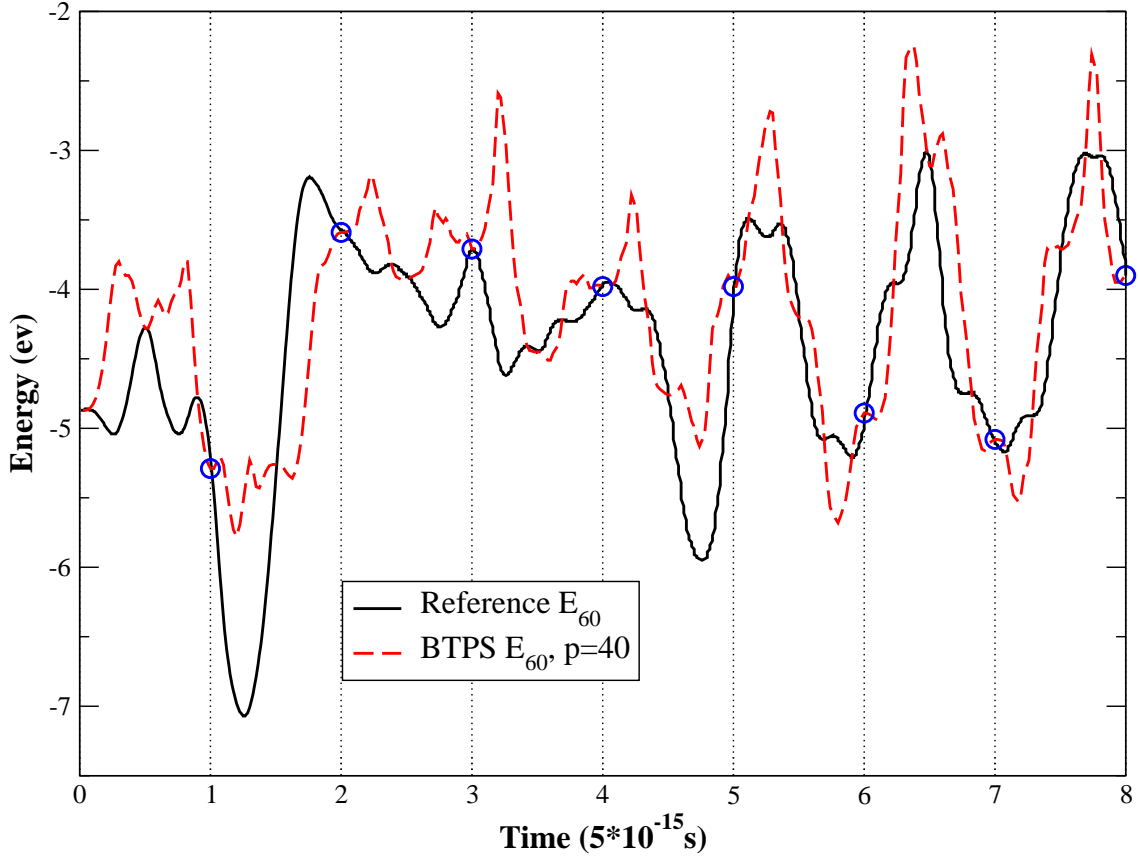}
  \caption{Evolution of the energy expectation for the HOMO level. BTPS can capture efficiently and exactly 
the reference solution at the end of each time interval using $p=40$ points for the rectangle quadrature rule
 (the BTPS solutions are pointed out in the plot using the circle symbols). 
We note that the intermediate solutions obtained with BTPS have no physical meaning.
}
  \label{fig3}
\end{figure}

Finally, similarly to the Gaussian propagation scheme, $p=40$ integration points suffice for the BTPS approach 
to accurately obtain all the solutions at the end of each time period. 
Using BTPS the numerical integration is performed on a projected space, and this operation does 
not require high-order integration schemes as discussed in Section \ref{secIIIc}. 
As compared to direct  or Gaussian propagation schemes, 
BTPS reduces drastically the computational costs meanwhile stable and accurate.
However and as shown in Figure \ref{fig3}, intermediate solutions that can be computed within the intervals
have no physical meaning. It should be noted that the same remark applied to the Gaussian scheme, although the 
(unphysical) solutions
within intervals were not represented in Figures \ref{fig1} and \ref{fig2} for clarity.

\begin{figure}
\centering
\includegraphics[width=\columnwidth]{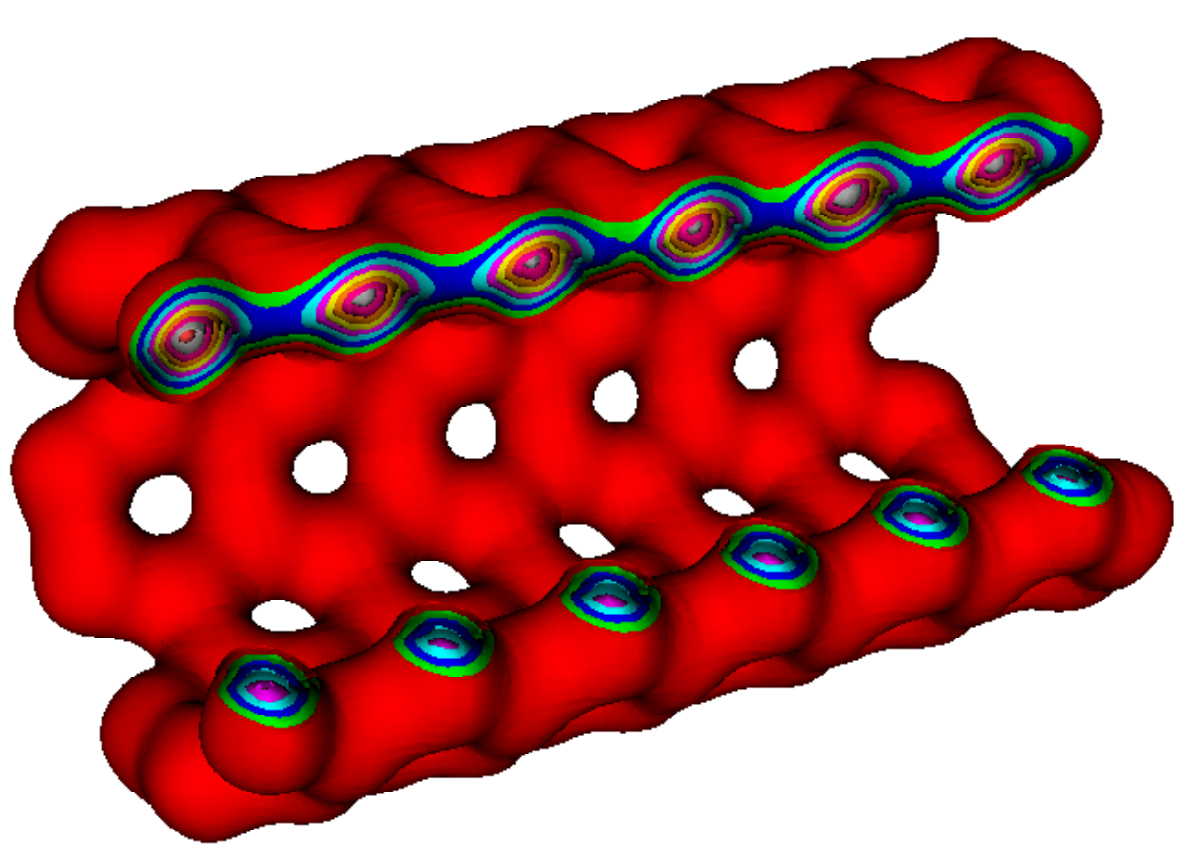}
\caption{Contour plot showing the 3D electron density at \mbox{$t=0$}. }
\label{fig4}
\end{figure}

\begin{figure}
\centering
\includegraphics[width=\columnwidth]{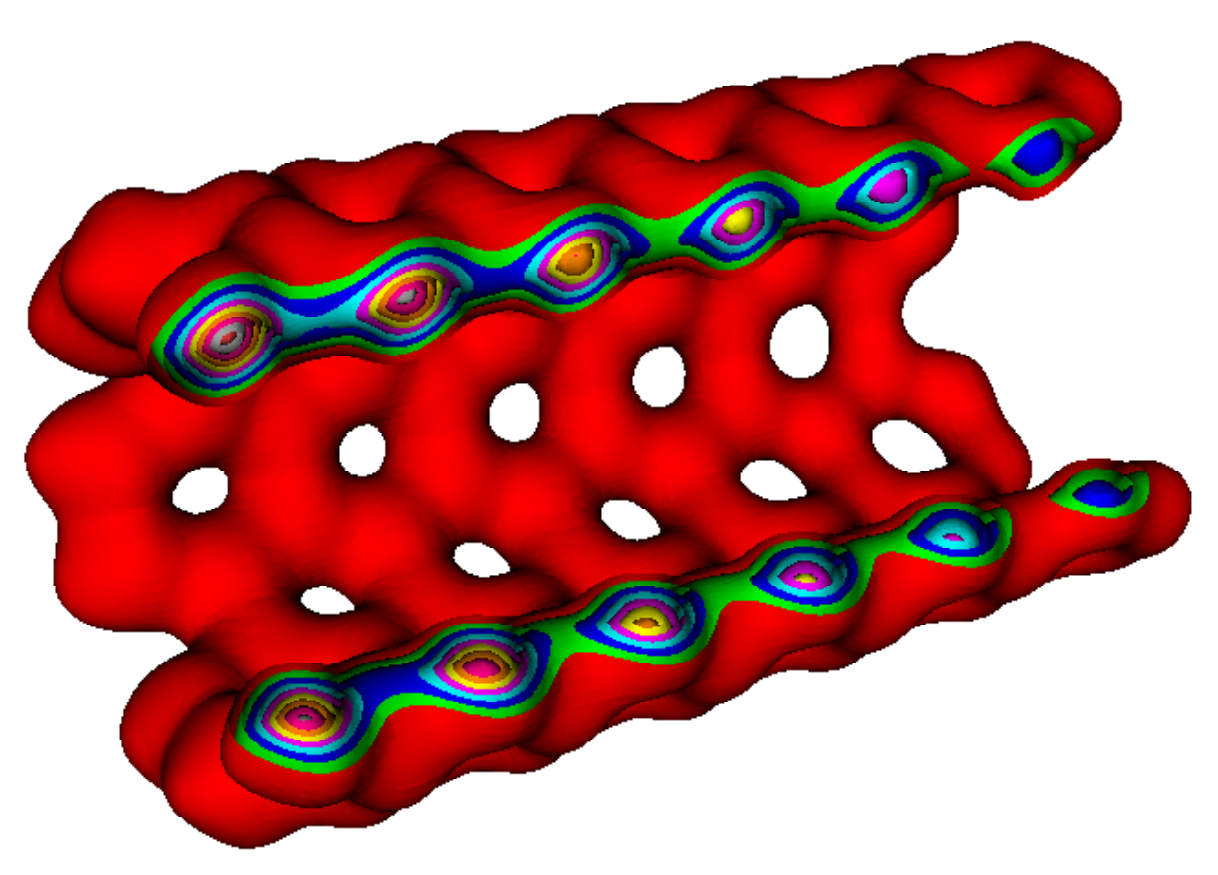}
\caption{Contour plot showing the 3D electron density at \mbox{$t=T/4$}, electrons have moved to the left edge.}
\label{fig5}
\end{figure}

From the solutions on the wave functions, one can now investigate many properties of the CNT. 
In particular, within the real-space mesh framework,  the  electron density is given by:
\begin{equation}
{\mathbf{n}(t)}=\sum_{j=1}^{N_e}|\mbox{\boldmath $\psi$}_j(t)|^2.
\end{equation} 
As an example, the results of the 3D simulations on the electron density are represented  
in Figures \ref{fig4} and \ref{fig5} respectively at $t=0$ and $t=T/4$ using a contour plot.


\begin{figure}
  \centering
  \includegraphics[angle=-90,width= \columnwidth]{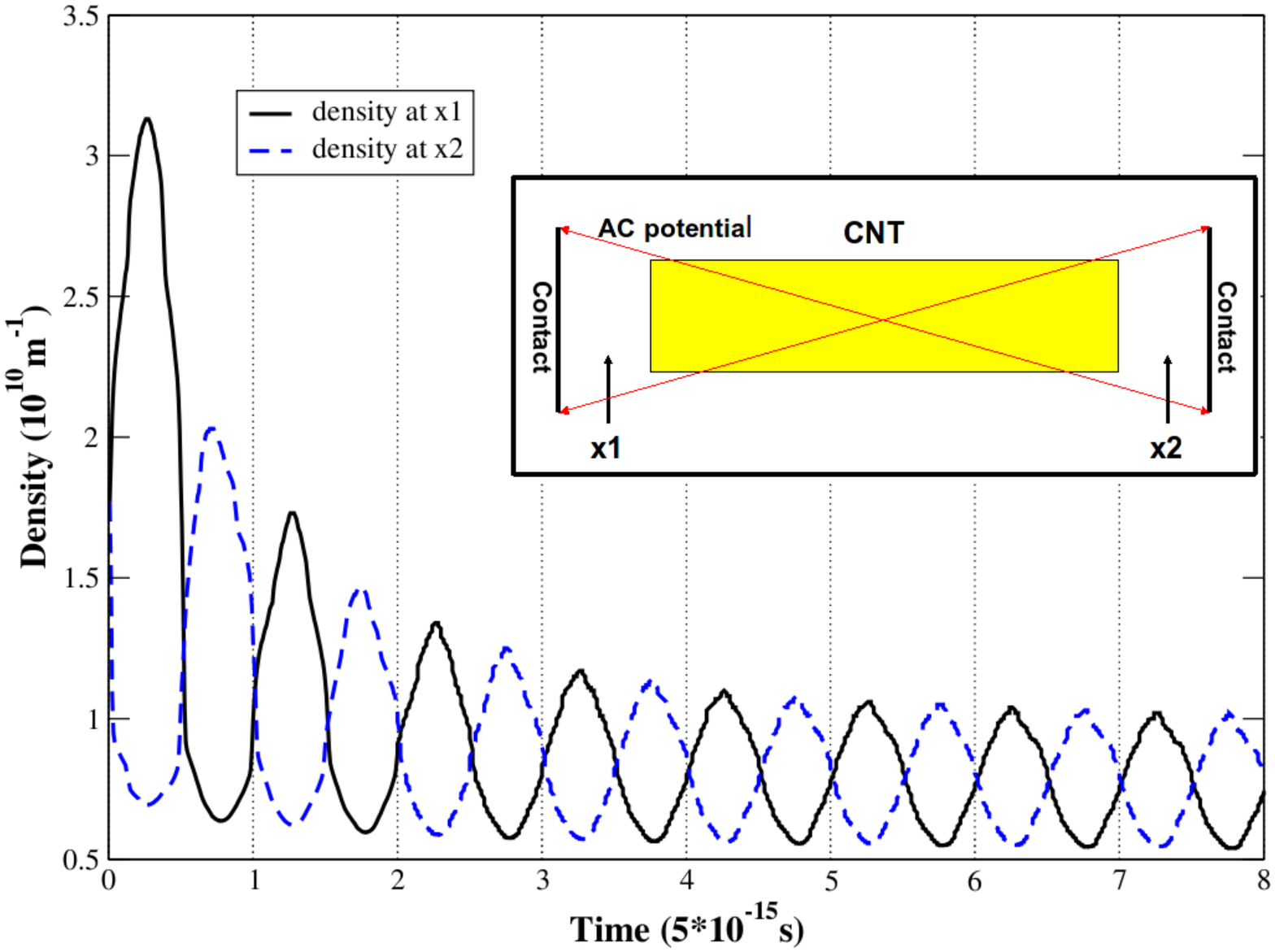}
  \caption{Evolution of the 1D projection of the electron density at two given positions.
In this example, the positions $x_1$and $x_2$ are selected between contacts and edges of the CNT, and
the results are then representative of the CNT field emission.
}
  \label{fig6}
\end{figure}

In order to better illustrate the time evolution of the density, it is possible to calculate
the variation of the 1D projection of the electron density on the longitudinal axis 
i.e. $n_{1D}(x)=\int_{y,z}n(x,y,z,t) dydz$. 
Figure \ref{fig6} shows how this 1D electron density calculated
using the direct propagation scheme and
at some particular positions along $x$, evolves over time. 
It should be then noted that much fewer points that the $p=120$ required by $\Delta_t$ using the direct approach,
could obviously be sufficient to accurately capture the variation on the density. Indeed, the electron density as well
as  other integrated quantities, are likely to exhibit much weaker variations as compared to the
variations of the individual wave functions.
Let us then suppose that the variation of the 
electron density can be captured using (at most) $30$ points by $\Delta_t=T$, BTPS
could then be efficiently used to compute these intermediate 
solutions using a new time interval of $\Delta_t=T/30$. 
For the same degree of accuracy, indeed, the direct propagation scheme would
 now require solving $4$ ($120/30$) large  eigenvalue problems within the new $\Delta_t$, but
BTPS would still require solving only one (and $40/30\simeq 1$ small reduced eigenvalue problem).


\section{Conclusion}\label{secV}

In this work, we have investigated three different spectral-based propagation techniques 
for time dependent quantum system: Direct, Gaussian and BTPS approaches.
These numerical schemes have been applied  to study 
 the AC response of an isolated single wall carbon nanotube using a real-space mesh techniques framework, 
empirical pseudopotential and non-interacting TDDFT calculations. 
Using the direct approach, the time-ordered evolution operator is solved 
via a step-by-step diagonalization procedure of the time-dependent Hamiltonian.
Spectral-based schemes are known as robust and accurate but are
traditionally considered too computationally expensive for addressing large systems.
The new eigenvalue solver FEAST, however, can efficiently address these problems and 
 provide performances and scalability.
We have also pointed out that two numerical errors do appear in 
time propagation problems: a quadrature error resulting from the integral on the Hamiltonian,
and an error resulting from the decomposition of the exponential. In contrast to the more
conventional direct approach, our proposed Gaussian scheme demonstrates that it is possible to
obtain the solutions at the end of each time interval using
a reduced number of quadrature points (i.e. reduction of the quadrature
 error using a higher-order integration scheme), while preserving accurate exponential decompositions.
Finally the BTPS scheme reduces
not only the number of eigenvalue problems to solve by intervals, but also the size of each eigenvalue problem.
Since only one large scale eigenvalue problem and a small number of reduced eigenvalue problems need to 
be solved by time intervals, the computational efficiency and time savings offer by BTPS are significant.

It is straightforward to note that the direct scheme can also be used to address self-consistent TDDFT calculations 
 using the adiabatic local density approximation (ALDA) where the potentials $v_H$ and $v_{xc}$ (\ref{eqvks}) 
need to be calculated at each time step in function of the local density. 
In order to take advantage of the BTPS scheme for interacting systems, however, one 
would need to adequately   
keep track of the variation of the density self-consistently with the local potential
by considering a small enough time interval $\Delta_t$. 
Since  density and potential are likely to exhibit much weaker variations 
with time as compared to individual wave functions,  BTPS should still provide significant 
computational time savings as compared to the direct approach.

To summarize, the spectral-based propagation schemes proposed here with in particular the optimized BTPS approach, 
are potentially capable to open new perspectives in time dependent simulations of large-scale quantum systems.
Possible applications of these techniques range from obtaining  
accurately   the excited states of arbitrary molecules and nanostructures,
to efficient characterization of high frequency responses of emerging nanoelectronic materials and devices.

\begin{acknowledgments}
  The authors wish to acknowledge helpful discussions
with Dr. Sigfrid Yngvesson. This material is based upon work supported by the National Science Foundation: Grants
No. ECCS 0725613, and No ECCS 0846457.
\end{acknowledgments}

\bibliography{cp2010}

\end{document}